# Empirical Comparison of Envelope-Tracking and Time-Domain Adaptive Integral Methods for Surface Integral Equations

Guneet Kaur and Ali E. Yılmaz

## Abstract

This paper presents a detailed evaluation of the envelope-tracking adaptive integral method (ET-AIM), an FFT-accelerated algorithm for analyzing electromagnetic scattering. ET-AIM is used to solve progressively more complex benchmark scattering problems and key parameters of the method (the auxiliary grid size, near-zone size, temporal basis function type, time-step size, and iterative solver tolerance) are optimized. The computational costs and accuracy of ET-AIM are compared to its time-domain counterpart, the time-domain adaptive integral method (TD-AIM), in the high-frequency regime, where the spatial discretization of the scattering object is determined by the minimum wavelength of interest rather than its geometrical features. Numerical results show that although ET-AIM and TD-AIM computation times are comparable when the bandwidth of interest is wide, the ET-AIM marching costs are dominated by iterative solution rather than scattered-field computations ("right-hand-side" computations) and that as the bandwidth of interest becomes narrower than $\mp 50\%$ of the center frequency, ET-AIM computational costs become significantly smaller than TD-AIM ones. ET-AIM is also shown to efficiently solve large and complex scattering problems whose solution by TD-AIM is impractical.

# 1 INTRODUCTION

Perhaps the most efficient approach for transient scattering analysis, especially for finding early-time or non-linear responses from piecewise homogeneous objects, is to formulate surface integral equations and solve them for unknown currents by using the time-domain marching-on-in-time (TD-MOT) method. This approach becomes even more potent when accelerated by fast algorithms like the plane-wave time-domain method [1] or the time-domain adaptive integral method (TD-AIM) [2]. Time-domain methods lose efficiency, however, whenever the maximum frequency of interest is (much) larger than the bandwidth of interest; this is because they must use time steps whose sizes are inversely proportional to the maximum frequency rather than the bandwidth of interest [7].

An attractive approach for improving integral-equation methods for transient scattering analysis is to use "envelope tracking," where surface integral equations are formulated and solved for the *complex envelopes of unknown currents* by time marching [3]-[7]. The complex envelope of a band-pass signal is its baseband representation [8]; thus, the sampling rate needed to resolve a signal's complex envelope is dictated not by its maximum frequency content but by its bandwidth, i.e., envelope-tracking methods must use time steps whose sizes are inversely proportional to the bandwidth. As a result, envelope-tracking methods solve for a (much) smaller number of temporal degrees of freedom compared to traditional time-domain methods for narrow-band problems. Despite this improvement, the envelope-tracking marching-on-in-time (ET-MOT) method has high computational costs [3],[5]: It requires $O(N_s^2)$ seconds to fill impedance matrices and $O(N_s^2 + N_g N_s)$ bytes to store these impedance matrices and the current-envelope samples found at previous time steps while marching in time; here, $N_s$ and $N_t$ are the number of spatial and temporal degrees of freedom and $N_g$ is determined by the maximum number of time steps that the fields require to traverse across two points on the scattering surface [2],[5]. In addition, ET-MOT requires $O(N_t N_s^2 + N_t \bar{N}_I N_0)$ seconds for time marching; this "marching time" consists of $O(N_t N_s^2)$ seconds spent for the "right-hand-side (RHS) time" of computing fields radiated by currents at previous time steps and $O(N_t \bar{N}_I N_0)$ seconds spent for the "iterative solution time" of finding the current-envelope samples at the present time step, where the iterative solver requires on average $\bar{N}_I$ iterations to converge at each time step and the matrix representing the linear equations being solved has $N_0$ non-zero entries.

Recently, the envelope-tracking adaptive integral method (ET-AIM) was proposed to reduce the computational complexity of the ET-MOT procedure [6],[7]. ET-AIM, just like its frequency- and time-domain counterparts [9],[2], uses an auxiliary regular grid to enclose the scattering object and exploits the space-time translational invariance of the Green function with 3-D space and 4-D space-time FFTs [7]. As a result, ET-AIM requires $O(N_C + N_s)$ seconds to fill the necessary matrices, $O(N_g[N_C + N_s])$ bytes to store the unique entries of these matrices and the current-envelope samples found at previous time steps, $O(N_t N_C [\log N_C + \log^2 N_g])$ seconds to compute delayed fields, and $O(N_\text{t} \bar{N}_\text{I} N_\text{C} \log N_\text{C})$ seconds to iteratively solve for the current-envelope samples. Here, $N_\text{C}$ is the number of nodes on the auxiliary grid.

This paper compares the performance of ET-AIM to its time-domain counterpart TD-AIM. Theoretical comparison of the two methods is limited because of two reasons: (i) *The comparison depends on the frequency regime of analysis*. While the comparison is rather straightforward in the low-frequency regime, where the spatial discretization lengths are dictated by geometrical details of the scattering object, it is complicated in the high-frequency regime, where the spatial discretization lengths are determined by the minimum wavelength of interest. (ii) *The methods have different accuracy-efficiency tradeoffs*. When identical spatial and temporal basis functions are used, envelope-tracking methods generally yield more accurate results than their time-domain counterparts because of smaller integration and interpolation errors [10]. Therefore, for a desired level of accuracy, the computational costs of envelope-tracking methods can be reduced by using different discretization parameters. Owing to these two factors, only limited deductions regarding ET-AIM and TD-AIM performances can be made theoretically in the high-frequency regime. To be able to compare these methods systematically, this paper adopts an empirical approach: Two simulators based on ET-AIM and TD-AIM are executed on the same computer to solve several benchmark scattering problems, the errors made by the methods are constrained, their parameters are chosen to minimize the computational costs, and the marching times of the simulators are measured. The results are contrasted over a range of bandwidths and problem sizes.

ET-AIM is also compared to TD-AIM in [7], where the scattering problems solved and the parameters used are identical to those in this paper. This paper is different from [7] in three aspects: (i) It presents a more in-depth empirical comparison of ET-AIM and TD-AIM. Specifically, unlike in [7], the two components of the marching cost are measured and analyzed separately here; the results highlight the fact that the marching times of envelope-tracking and time-domain methods are distributed very differently between the RHS time and the iterative solution time. (ii) It employs a frequency-domain error norm in addition to the time-domain error norm of [7]; as a result, ET-AIM errors are shown to behave differently than TD-AIM errors as a function of frequency. (iii) It lists the full set of optimal parameters used in both papers and facilitates repeatability of the results.

The rest of the paper is organized as follows: Section 2 describes the methodology adopted for the empirical comparison. Section 3 presents ET-MOT and compares the accuracy of envelope-tracking and time-domain methods. Section 4 briefly presents ET-AIM, its computational complexity, and its theoretical comparison to TD-AIM. Section 5 presents the empirical comparison of ET-AIM and TD-AIM for benchmark scattering problems. Section 6 presents numerical results that demonstrate ET-AIM's generality for solving complex scattering problems. Section 7 presents the conclusions.

## 2 EMPIRICAL COMPARISON METHODOLOGY

Various scattering problems are solved in this paper to compare different methods; this section details the problems solved and the comparison methodology.

### 2.1 Scattering Objects and Integral Equations

In all problems solved, the scattering object is assumed to be perfect electrically conducting and to reside in a homogeneous medium with permittivity $\varepsilon$ and permeability $\mu$. For the envelope-tracking solution, a combined field integral equation (CFIE) is formulated in terms of the complex envelopes of the incident fields $\{\tilde{\mathbf{E}}^{\text{inc}}, \tilde{\mathbf{H}}^{\text{inc}}\}$, scattered fields $\{\tilde{\mathbf{E}}^{\text{sca}}, \tilde{\mathbf{H}}^{\text{sca}}\}$, and induced current density $\tilde{\mathbf{J}}$; the derivation is detailed in [7]. The parameter used to linearly combine the electric- and magnetic field integral equation (EFIE and MFIE) to obtain the CFIE is denoted by $\alpha$, which is between 0 and 1 [7]. For the time-domain solution, the standard CFIE is formulated in terms of the incident fields $\{\mathbf{E}^{\text{inc}}, \mathbf{H}^{\text{inc}}\}$, scattered fields $\{\mathbf{E}^{\text{sca}}, \mathbf{H}^{\text{sca}}\}$ and the induced current density $\mathbf{J}$ [2]. Here and throughout the paper, a tilde above a function indicates that it is the complex envelope of the underlying function.

### 2.2 Excitation

Each object of interest is illuminated by a cosine modulated Gaussian plane wave, i.e.,

$$\begin{Bmatrix} \mathbf{E}^{\text{inc}}(\mathbf{r},t) \\ \mathbf{H}^{\text{inc}}(\mathbf{r},t) \end{Bmatrix} = \begin{Bmatrix} \hat{p} \\ \hat{k} \times \hat{p}/\eta \end{Bmatrix} e^{-\frac{(t-\mathbf{r}\cdot\hat{k}/c-t_{\text{d}})^2}{2\sigma^2}} \cos(2\pi f_{\text{c}}[t-\mathbf{r}\cdot\hat{k}/c-t_{\text{d}}]) \qquad (1)$$

Here, $\hat{p}$ is the unit polarization vector, $\hat{k}$ is the unit vector in the direction of the wave propagation, $\eta = \sqrt{\mu/\varepsilon}$ is the intrinsic impedance of the surrounding medium, $c = 1/\sqrt{\mu\varepsilon}$ is the speed of light in this medium, $f_{\text{c}}$ is the center frequency, and the Guassian pulse has the standard deviation $\sigma$ and is shifted in time by $t_{\text{d}}$. Less than $\sim 2.2 \times 10^{-3}\%$ of the energy of this pulse is outside the time interval $t_{\text{d}} \pm 3\sigma$ and frequency band $f_{\text{c}} \pm 3/2\pi\sigma$.

### 2.3 Time and Frequency Widths

In the following, all fields are assumed to be *essentially* time- and band-limited, i.e., the fields are assumed to be vanishingly small on the scattering surface for times $t \leq 0$ and $t \geq T^{\text{s}}$ and frequencies $f < f_{\text{min}}$ and $f > f_{\text{max}}$; therefore, the essential time- and band-width of the fields are defined as $t_{\text{tw}} = T^{\text{s}}$ and $f_{\text{bw}} = (f_{\text{max}} - f_{\text{min}})/2$. The narrowness of the frequency band of interest is measured using the ratio $\chi = f_{\text{max}}/f_{\text{bw}}$; e.g., $\chi = 1$ for baseband analysis ($f_{\text{min}} = -f_{\text{max}}$), $\chi = 2$ when the bandwidth covers the full (positive) frequency spectrum from $f_{\text{min}} = 0$ to $f_{\text{max}}$, $\chi = 2.22...$ for a 10:1 (maximum-to-minimum frequency ratio) bandwidth, $\chi = 3$ for a $\mp 50\%$ band around the center frequency, $\chi = 11$ for a $\mp 10\%$ band around the center frequency, $\chi = 101$ for a $\mp 1\%$ band around the center frequency, and $\chi \to \infty$ for an un-modulated (time-unlimited) sinusoid. The parameters $T^s$, $f_{\min}$, and $f_{\max}$ are generally not known *a priori* because the essential time- and band-width of the total (incident plus scattered) fields can be different from the incident fields and because the cut-off values beyond which the fields are considered to vanish depend on various factors that are unknown initially and can be fully determined only after the analysis is performed [7]; e.g., the essential time-width for the analysis can be reduced if less accurate results are acceptable in the computed radar cross section or increased if more accurate results are desired.

## 2.4 Error Measure

To quantify the accuracy of the different solutions, the co-polarized bistatic radar cross section (RCS) $\sigma_{\theta\theta}(f,\theta,\phi)$ is computed by post-processing the solution and the error in the RCS is measured using the dB-error norm

$$err^{TH}_{\theta\theta,\mathrm{dB}}(f) = \frac{\int_0^{2\pi}\int_0^{\pi}[\sigma^{TH}_{\theta\theta}(f,\theta,\phi) - \sigma^{TH,\mathrm{ref}}_{\theta\theta}(f,\theta,\phi)]^2 \sin\theta d\theta d\phi}{\int_0^{2\pi}\int_0^{\pi}[\sigma^{TH,\mathrm{ref}}_{\theta\theta}(f,\theta,\phi)]^2 \sin\theta d\theta d\phi} \qquad (2)$$

where $\sigma^{TH}_{\theta\theta} = \max(10\log[\sigma_{\theta\theta}(f,\theta,\phi)], TH) - TH$ is the bistatic RCS value filtered and adjusted such that only values larger than the threshold $TH$ remain and the RCS values at $TH$ are set to $0\ \mathrm{dB}$. The $TH$ value is arbitrary and can be chosen to emphasize or deemphasize some features of the errors in the computed RCS. Throughout the paper, the threshold value $TH$ is set to be $80\ \mathrm{dB}$ lower than the peak value of $\sigma_{\theta\theta}$ at the frequency of interest.

Though there exist many other error norms that can be used to measure the accuracy of numerical methods, e.g., L2 error norm of RCS [2],[11], the electric field [2], the induced current density [10], etc., the above dB-error norm is introduced here because it also captures errors in values (much) smaller than the peak value of RCS, which are essentially invisible when linear error norms are used. The dB error norm de-emphasizes the errors in the large RCS values compared to linear error norms as a result of this operation; nevertheless, it remains more sensitive to the larger RCS values compared to smaller ones because the larger is the difference between the RCS and the threshold, the larger (more important) is its contribution to the error. Compared to linear error norms, the dB-error norm in (2) will yield values that better reflect the visually observed differences in logarithmic RCS plots. These properties are also shared by the dB-error norm used in [7]. In contrast to [7], where the error is computed using the backscattered fields (single direction) with respect to time (entire time-width of interest), the dB-error norm in (2) is computed using the RCS (single frequency) with respect to the observation angle (entire solid angle).

## 2.5 Computational Cost Measurements

Various timing and memory measurements are reported in this paper and in [7]. The results in both papers were obtained on Lonestar 4 at the Texas Advanced Computing Center, which is a Linux cluster comprised of $3.33\ \mathrm{GHz}$

6-core Xeon processors [12]. The ET-AIM, TD-AIM, and FD-AIM in [7] and here were parallelized using an MPI based implementation of the 1-D slab decomposition based algorithm described in [2],[13]. The computations were performed using the minimum number of processors dictated by the memory constraints and activating a single core in each processor. The reported 'serialized' computational costs in [7] and in this paper were obtained by measuring the wall-clock time and the peak memory required among processors and multiplying these with $P$, the total number of cores used.

## 3 ENVELOPE-TRACKING MARCHING-ON-IN-TIME (ET-MOT) METHOD

In order to solve the CFIE, the complex envelope of the surface current density on the scattering object is discretized using $N_s N_t$ space-time basis functions:

$$\tilde{\mathbf{J}}(\mathbf{r},t) \cong \sum_{k'=1}^{N_s}\sum_{l'=1}^{N_t}\tilde{\mathbf{I}}_{l'}[k']\mathbf{S}_{k'}(\mathbf{r})T(t-l'\Delta t) \qquad (3)$$

Here, $\tilde{\mathbf{I}}_{l'}$ is the vector of unknown current-envelope coefficients at time $l'\Delta t$, $\mathbf{S}_{k'}$ is the $k'^{\text{th}}$ RWG basis function defined on a triangular mesh of the scattering surface [14], $\Delta t = \beta/f_{\text{bw}}$ is the time step size, $1/2\beta$ is the oversampling over the Nyquist rate, $N_t = \lfloor T^s/\Delta t \rfloor$ is the total number of time steps, and $T$ is the temporal basis function, which is either a band-limited interpolatory function (BLIF) [10],[15] or a causal piecewise polynomial interpolatory function (CPPIF) [10],[16]. After the current density is discretized, Galerkin testing is performed in space at times $\Delta t, 2\Delta t,..., N_t \Delta t$, which results in the following system of equations (in case BLIFs are used, an additional extrapolation step is required to express the resulting system of equations in a causal form similar to the following equations [15]):

$$\tilde{\mathbf{Z}}_0 \tilde{\mathbf{I}}_l = \tilde{\mathbf{V}}_l^{\text{inc}} - \sum_{l'=\max(1,l-N_g)}^{l-1}\tilde{\mathbf{Z}}_{l-l'}\tilde{\mathbf{I}}_{l'} \text{ for } l = 1,2,...,N_t \qquad (4)$$

In (4), $N_g = \lfloor R^{\max}/c\Delta t + l^{\max} \rfloor$, where $R^{\max}$ is the maximum distance between the two points on the scattering surface and $l^{\max}$ is the causal length of $T$ [7]. The entries of the impedance matrices $\tilde{\mathbf{Z}}_0,...,\tilde{\mathbf{Z}}_{N_g}$ and the excitation vector $\tilde{\mathbf{V}}_l^{\text{inc}}$ are given in [5]; the $\tilde{\mathbf{Z}}_0$ matrix, also referred to as the "immediate-interaction matrix," gives the contribution to the envelope of scattered fields at a given time from the current envelope samples at the same time.

The major computational costs of this ET-MOT scheme are the time needed to fill the matrices ("matrix fill time"), the memory required to store the impedance matrices and the current vectors ("memory requirement"), the time needed to form the right-hand-side of (4) ("the RHS time"), and time needed to iteratively solve of (4) ("iterative solution time"), which scale as $O(N_s^2)$, $O(N_s^2)$, $O(N_s^2)$, and $O(\bar{N}_I N_0)$ per time step, respectively. As mentioned in the Introduction, $\bar{N}_I$ is the average iteration count per time step for iterative solver convergence and $N_0$ is the number of non-zero entries in $\tilde{\mathbf{Z}}_0$.

## 3.1 Accuracy Comparison to Time-Domain Marching-on-in-Time (TD-MOT)

It was shown in [5] that the errors made by envelope-tracking and time-domain MOT solvers depend differently on the bandwidth of interest; specifically, it was shown that the ET-MOT errors decrease whereas TD-MOT ones stay constant as the bandwidth of interest narrows. As mentioned in [5], this is primarily because the numerical integration errors [5],[10] when computing the impedance matrix entries is lower in envelope-tracking solvers compared to time-domain solvers when identical temporal basis functions, oversampling rates, and standard numerical cubature rules are used. Here, it is shown that the errors in the methods also depend differently on the frequency; specifically, it is shown that the ET-MOT errors are more frequency dependent than TD-MOT ones. This is because the interpolation, extrapolation, and integration errors of the two methods are different [10].

To demonstrate this effect, scattering from a 2-m radius sphere was simulated. The sphere was illuminated by the cosine modulated Gaussian plane wave in (1) with a carrier frequency of $f_c = 200$ MHz using two different standard deviations: $\sigma = 3/200\pi$ $\mu$s and $\sigma = 3/4\pi$ $\mu$s, which correspond to essential bandwidths of $f_{bw} \approx 100$ MHz ($\chi \approx 3$) and $f_{bw} \approx 2$ MHz ($\chi \approx 100$), respectively; the time delay was set to $t_d = 8\sigma$ in each case. Identical CFIE combinations ($\alpha = 0.5$), meshes, spatial basis functions, cubature rules, singularity extraction techniques, temporal basis functions, oversampling rates, and iterative solver tolerances ($\text{err}^{\text{tol}} = 10^{-9}$) were used for ET-MOT and TD-MOT. The impact of oversampling rate on the results was investigated by using various time-step sizes: In TD-MOT simulations $\Delta t^{\text{TD}} = 0.67$ ns ($\beta^{\text{TD}} \approx 1/5$) or $\Delta t^{\text{TD}} = 0.22$ ns ($\beta^{\text{TD}} \approx 1/15$) were used[1]; in ET-MOT simulations $\Delta t = 2$ ns ($\beta \approx 1/5$) or $\Delta t = 0.67$ ns ($\beta \approx 1/15$) were used in the broad-band case and $\Delta t = 0.1$ $\mu$s ($\beta \approx 1/5$) or $\Delta t = 33$ ns ($\beta \approx 1/15$) were used in the narrow-band case. In the following, the results are compared to the frequency-domain method-of-moments solution of the CFIE; the frequency-domain solution used the same CFIE linear combination parameter, mesh, spatial basis functions, cubature rules, singularity extraction techniques, and iterative solver tolerance as ET-MOT and TD-MOT. The reference frequency-domain forward- and back-scattered RCS are plotted in Figs. 1(a)-(b) for the two different frequency bands that were simulated.

*3.1.1 Causal Piecewise Polynomial Interpolatory Functions (CPPIFs)*

The ET-MOT and TD-MOT errors when using CPPIFs are shown as a function of frequency in Figs. 1(c)-(d). It is observed in Fig. 1(c) that ET-MOT yields very large errors for the broad-band ($\chi \approx 3$) simulation. This is because the system of equations in (4) obtained using CPPIFs loses diagonal dominance as the bandwidth widens, which results in unstable solutions. In contrast, Fig. 1(d) shows that ET-MOT yields much more accurate results for the narrow-band ($\chi \approx 100$) simulation, with the best accuracy achieved at the center frequency, where the errors are on the order of iterative solver tolerance $\text{err}^{\text{tol}}$ and the results are essentially indistinguishable from those obtained by the frequency-domain method-of-moments solution.

---
[1]Throughout the paper a superscript TD over a variable indicates that it is the time-domain counterpart of a variable that was originally defined for envelope-tracking.

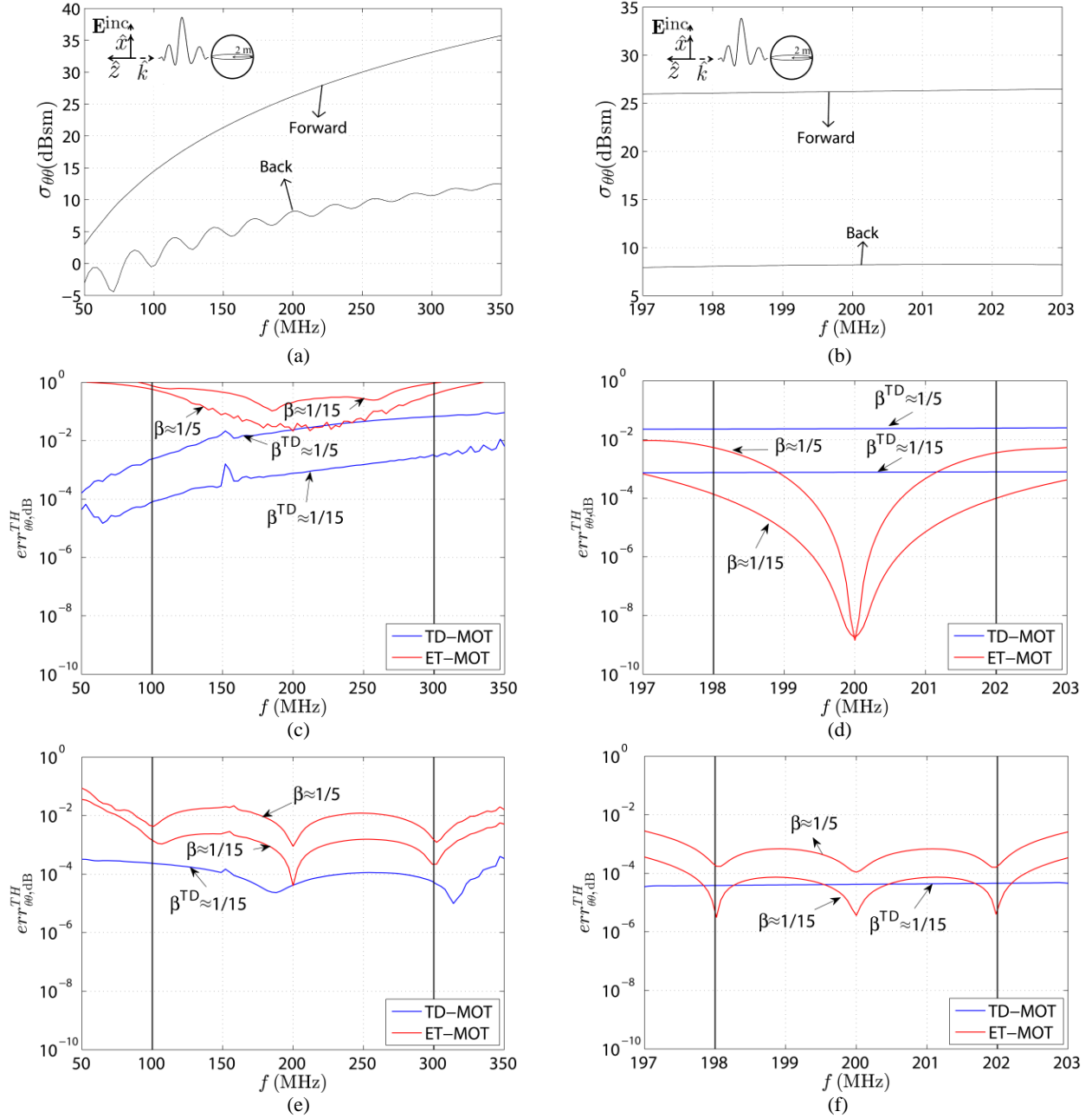

Figure 1: ET-MOT vs. TD-MOT for a 2-m radius PEC sphere. Top row: Reference RCS data found from the frequency-domain method-of-moments solution in the forward- and back-scattering directions. Middle row: Errors when CPPIFs are used as temporal basis functions. Bottom row: Errors when BLIFs are used as temporal basis functions. Solid black lines show the essential bandwidth of the incident Gaussian plane wave; a broad band ($\chi \approx 3$, left column) and a narrow band ($\chi \approx 100$, right column) simulation are shown. CPPIF of order $Q = 3$ and BLIF of half-width $M = 3$ were used for ET-MOT; CPPIF of order $Q = 4$ and BLIF of half-width $M = 5$ were used for TD-MOT. The error norm threshold was $TH = \max(10 \log \sigma_{\theta\theta}) - 80$ dB.

Compared to TD-MOT errors, which vary only slightly with respect to frequency (about one order of magnitude across the band) for the broad-band simulations and are essentially constant across the band for the narrow-band simulations, ET-MOT errors show larger variations with respect to frequency (when the results are stable). This is because CPPIFs make the largest interpolation errors at the highest frequency of interest and the lowest errors at 0 frequency. Thus, for band-limited analysis with ET-MOT, where the center frequency is shifted to the baseband, CPPIFs have extremely small interpolation errors near the center frequency; this increases the accuracy of ET-MOT near the center frequency. It is important to observe that at the upper limit of the band (highlighted with solid lines in Fig. 1) the ET-MOT and TD-MOT interpolation errors are identical for identical oversampling rates but the envelope-tracking solution is an order of magnitude more accurate than the time-domain one. This is because the larger time-step sizes used by ET-MOT relative to the spatial mesh size results in lower integration errors when standard cubature rules are used [5]. Fig. 1 also shows that decreasing the oversampling rate (increasing $\beta, \beta^{\text{TD}}$ from $\approx 1/15$ to $\approx 1/5$) results in 10-100 times higher errors for both ET-MOT (when stable) and TD-MOT. Note that, when ET-MOT solution is stable, the results are reasonably accurate (less than 1% error) in the entire band even for the smaller oversampling rate (Fig. 1(d)).

*3.1.2 Band-Limited Interpolatory Functions (BLIFs)*

The ET-MOT and TD-MOT errors when BLIFs are used are shown as a function of frequency in Figs. 1(e)-(f). Here, the extrapolation scheme described in [15] is used with $N_\omega = 2N_{\text{samp}}$ where $N_{\text{samp}}$ is set to 3 and 5 for ET-MOT and TD-MOT, respectively, as these values were found to optimize the extrapolation performance. Figs. 1(e)-(f) show that the ET-MOT error decreases as the bandwidth narrows and as the oversampling rate increases ($\beta$ decreases). In contrast, TD-MOT is unstable in both broad- and narrow-band cases for the small oversampling rate ($\beta^{\text{TD}} \approx 1/5$) and has errors that are relatively insensitive to the bandwidth of interest. Moreover, similar to the case of CPPIFs, ET-MOT errors are found to be frequency dependent whereas TD-MOT ones are essentially constant with respect to frequency. Increasing the order of interpolation by increasing half-width $M$ of the BLIF beyond 3 for ET-MOT and 5 for TD-MOT did not reduce the errors. This is because extrapolation errors, which are comparable for higher values of $M$, dominate interpolation errors. Indeed, BLIFs exhibit smaller interpolation errors compared to CPPIFs but incur additional frequency-dependent extrapolation errors [10].

## 3.2 Synopsis

Fig. 1 shows that ET-MOT is prone to instabilities for broad-band simulations when CPPIFs are used. Similarly, TD-MOT is found to exhibit late-time instabilities for narrow-band simulations when CPPIFs are used. BLIFs of half-width parameter $M = 3$ for ET-MOT and $M = 5$ for TD-MOT are found to yield optimal performance. It is observed that a smaller oversampling rate can be used in ET-MOT compared to TD-MOT without significantly compromising accuracy when CPPIFs are used and without encountering late-time instabilities when BLIFs are used. The results suggest that CPPIFs should be used for narrow-band ET-MOT simulations and BLIFs for broad-band ones, whereas CPPIFs should be used for broad-band TD-MOT simulations and BLIFs for narrow-band ones.

## 4 ENVELOPE-TRACKING ADAPTIVE INTEGRAL METHOD (ET-AIM)

To accelerate the ET-MOT scheme, the scattering object is enclosed by a regular 3-D grid composed of $N_C$ nodes. At each time step, the following operations are performed in the ET-AIM algorithm [7]: (i) Anterpolation: The samples of current envelope at previous time steps at each spatial basis function ($\tilde{\mathbf{I}}_{\max(1,l-N_g)}[k']\mathbf{S}_{k'},\ldots,\tilde{\mathbf{I}}_{l-1}[k']\mathbf{S}_{k'}$ for time step $l$) are represented by nearby point sources on the auxiliary grid such that the fields produced by the two sets of sources are similar beyond a certain distance. (ii) Propagation: The sources on the auxiliary grid are radiated and the envelopes of the potentials are found at all points on the auxiliary grid at the present and some of the future time steps ($l, l+1, \ldots, l+X$ for time step $l$, where $X$ varies from 0 to $N_g$ depending on the time step [2]). (iii) Interpolation: The scattered field envelope at the present time step due to the current-envelope samples at the previous time steps is computed on the primary mesh by linearly combining the potential envelopes at the grid nodes computed in stage (ii). (iv) Near-field correction: Because the sources on the auxiliary grid do not reproduce fields accurately near the spatial basis function they represent, the fields in the near zone are (pre-)corrected by removing the effect of the computations performed in stages (i)-(iii). (v) Iterative solution: The current envelope at the present time step is found iteratively. This requires the computation of the scattered field envelope at the same time step, i.e., the "immediate interactions", due to the current envelopes that are guessed by the iterative solver. This computation is performed in four stages similar to (i)-(iv), where the first stage involves anterpolation of the guessed current envelopes. In effect, ET-AIM approximates (4) as [7]:

$$\tilde{\mathbf{Z}}_0^{\text{near}} + \tilde{\mathbf{Z}}_0^{\text{FFT}} \ \tilde{\mathbf{I}}_l \approx \tilde{\mathbf{V}}_l^{\text{inc}} - \sum_{l'=\max(1,l-N_g)}^{l-1} \tilde{\mathbf{Z}}_{l-l'}^{\text{near}} + \tilde{\mathbf{Z}}_{l-l'}^{\text{FFT}} \ \tilde{\mathbf{I}}_{l'}, \tag{5}$$

Here, $\tilde{\mathbf{Z}}_{l-l'}^{\text{near}}$ are (pre-)corrected impedance matrices, whose non-zero entries correspond to interactions between basis-testing function pairs in the near-zone of each other, and $\tilde{\mathbf{Z}}_{l-l'}^{\text{FFT}}$ are matrices that represent the interactions among all basis-testing pairs approximated by using the auxiliary grid. The entries of these matrices are given in [7]. Note that if BLIFs are used as temporal basis functions, then similar to ET-MOT, an extrapolation step is needed to obtain the final system of equations in a causal form as in (5).

The right-hand-side computations in (5) are accelerated using 4-D blocked space-time FFTs, as implemented for TD-AIM [2]. Similarly, the matrix-vector multiplications at each iteration that involve $\tilde{\mathbf{Z}}_0^{\text{FFT}}$ are accelerated using

3-D space FFTs just like the scheme in [17] for extending TD-AIM to the low-frequency regime. It is imperative to observe that there is a significant difference between ET-AIM and TD-AIM in the high-frequency regime: The TD-AIM near-zone size can generally be chosen large enough that all the immediate interactions are within the near zone and thus are computed without any AIM approximations—here it is assumed that CPPIFs are used as in [2]; when BLIFs are used, often the TD-AIM near-zone cannot be chosen large enough to contain all immediate (extrapolated) interactions and TD-AIM should also employ 3-D space FFTs to accelerate the iterative solution.

## 4.1 Theoretical Comparison to TD-AIM

Next, ET-AIM and TD-AIM computational costs are compared theoretically. In the following analysis, it is assumed that identical temporal basis functions and sampling rates are used in the two methods, i.e., the time step sizes are set to $\Delta t = \beta/f_{\text{bw}}$ for ET-AIM and $\Delta t^{\text{TD}} = \beta^{\text{TD}}/f_{\text{max}}$ for TD-AIM, where $\beta = \beta^{\text{TD}}$ by assumption; thus ET-AIM time step size is assumed to be $\chi = f_{\text{max}}/f_{\text{bw}}$ times larger than the TD-AIM one. It is also assumed that the same spatial basis functions and auxiliary grids are used in ET-AIM and TD-AIM; thus, $N_{\text{s}}$ and $N_{\text{C}}$ are also assumed identical. Lastly, it is assumed that CPPIFs are used; the analysis can be modified to BLIFs by taking into account the 3-D FFT operations required to compute the immediate interactions for TD-AIM.

As with ET-MOT, the major computational costs of ET-AIM and TD-AIM are the time needed to compute the necessary matrices ("matrix-fill time"), storage of these matrices and the current-envelope vectors ("memory requirement"), and the computation of the right-hand side ("RHS time") and iterative solution ("iterative solution time") of (5) or a similar equation for TD-AIM. The sum of time spent on right-hand-side and iterative solution computations comprises the total marching time. Most of these computational costs are a function of the frequency regime of interest and therefore, much like their classical MOT counterparts, the performance of ET-AIM and TD-AIM is a function of the frequency regime of interest.

*4.1.1 Low-Frequency Regime*

In the low-frequency regime of analysis, where the spatial discretization lengths are dictated by the geometrical details, the comparison is straightforward: In this regime, $N_{\text{g}} \approx N_{\text{g}}^{\text{TD}} \sim 1$, i.e., the number of non-zero impedance matrices is a small constant for both ET-AIM and TD-AIM. As a result, the matrix-fill time scales as $O(N_{\text{C}} + N_{\text{s}})$ for both methods, the memory requirement scales as $O(N_{\text{C}} + N_{\text{s}})$ for both methods, the RHS time scales as $O(N_{\text{t}} N_{\text{C}} \log N_{\text{C}})$ for ET-AIM and $O(N_{\text{t}}^{\text{TD}} N_{\text{C}} \log N_{\text{C}})$ for TD-AIM, and the iterative solution time, which dominates the RHS time for both methods, scales as $O(N_{\text{t}} \bar{N}_{\text{I}} N_{\text{C}} \log N_{\text{C}})$ for ET-AIM and as $O(N_{\text{t}}^{\text{TD}} \bar{N}_{\text{I}}^{\text{TD}} N_{\text{C}} \log N_{\text{C}})$ for TD-AIM. Here, $\bar{N}_{\text{I}} \approx \bar{N}_{\text{I}}^{\text{TD}}$ because the immediate interaction matrix is dense for both methods; thus, the two methods have similar costs per time step and the ET-AIM gain in the marching time over TD-AIM is proportional to the reduction in the number of time steps, i.e., the ET-AIM marching time is $N_\text{t}^\text{TD}/N_\text{t} \sim \chi$ times faster than the TD-AIM one.

*4.1.2 High-Frequency Regime*

In the high-frequency regime of analysis, where the object of interest is devoid of any sub-wavelength geometrical details and the spatial discretization lengths are comparable to the minimum wavelength of interest or the light-step size, $c\Delta t^\text{TD}$, the analysis is more complicated. This is because the larger time-step size used in envelope-tracking methods affects the number, sparsity, and conditioning of the impedance matrices.

The comparison remains straightforward for the matrix-fill time and memory requirement: The matrix-fill time scales as $O(N_\text{C} + N_\text{s})$ for both methods and the memory requirement scales as $O(N_\text{g}[N_\text{C} + N_\text{s}])$ for ET-AIM and as $O(N_\text{g}^\text{TD}[N_\text{C} + N_\text{s}])$ for TD-AIM. Typically, $N_\text{g}^\text{TD} \sim N_\text{s}^{1/2}$ and $N_\text{g} \sim \max(1, N_\text{s}^{1/2}/\chi)$; therefore, the ET-AIM memory requirement is lower than the TD-AIM one by a factor $\sim \min(N_\text{g}^\text{TD}, \chi)$ in the high-frequency regime [5].

The total ET-AIM and TD-AIM RHS times scale as $O\left(N_\text{t} N_\text{C} \log N_\text{C} + \log^2 N_\text{g}\right)$ and $O\left(N_\text{t}^\text{TD} N_\text{C} \times \log N_\text{C} + \log^2 N_\text{g}^\text{TD}\right)$, respectively. If the same time interval is simulated, i.e., $T^\text{s} = N_\text{t} \Delta t = N_\text{t}^\text{TD} \Delta t^\text{TD}$, then (i) the total number of time steps for ET-AIM is $\chi$ times smaller than that for TD-AIM, i.e., $N_\text{t} = N_\text{t}^\text{TD}/\chi$, and (ii) the number of ET-AIM impedance matrices is smaller by a factor of $N_\text{g}^\text{TD}/N_\text{g}$. These two facts imply that the right-hand-side computations are less costly for ET-AIM compared to TD-AIM. The iterative solution, however, is more costly for ET-AIM because all ET-AIM impedance matrices, including $\tilde{\mathbf{Z}}_0$, are denser compared to the corresponding TD-AIM ones. Indeed, the iterative solution time typically dominates the total marching time for ET-AIM, whereas it is a small portion of the total marching time for TD-AIM.[2] On the one hand, the gain in RHS time for ET-AIM is greater than $\chi$; on the other hand, the gain in the iterative solution time is less than $\chi$. Thus, the overall gain in the marching time is a complex function of $\chi$ and it is not clear if and by how much ET-AIM is faster than TD-AIM for a given value of $\chi$.

To estimate the value of $\chi$ where the ET-AIM marching time, which scales as $O(N_\text{t}[\bar{N}_\text{I} N_\text{C} \log N_\text{C} + N_\text{C} \times (\log N_\text{C} + \log^2 N_\text{g})])$ becomes less than the TD-AIM one, which scales as $O(N_\text{t}^\text{TD}[\bar{N}_\text{I}^\text{TD} N_0^\text{TD} + N_\text{C} \times (\log N_\text{C} + \log^2 N_\text{g}^\text{TD})])$, the following assumption is made in addition to those at the beginning of this section: The TD-AIM marching time is assumed to be approximated well as $O\left(N_\text{t}^\text{TD} N_\text{C} \log^2 N_\text{g}^\text{TD}\right)$; this is a valid simplification if the iterative solution time is negligible compared to the RHS time and if the object is large enough such that $\log^2 N_g^{TD} \gg \log N_C$ --or equivalently if $\log N_C \gg 4$ (if $N_g^{TD} \sim N_C^{1/2}$) or $\log N_C \gg 9$ (if $N_g^{TD} \sim N_C^{1/3}$). Under this condition, ET-AIM marching time will be faster than the TD-AIM one whenever $\chi \gg [\bar{N}_I \log N_C + \log^2 N_g]/\log^2 N_g^{TD}$, which holds true whenever $\chi \gg 1 + \bar{N}_I \log N_C/\log^2 N_g^{TD} \sim 1 + \bar{N}_I/\log N_C$ because $N_g < N_g^{TD}$ in general. If $\bar{N}_I$ does not change significantly with problem size, then the cross-over value of $\chi$ where ET-AIM marching time is faster than TD-AIM one decreases as the problem size increases. In other words, the larger the problem size, the larger the bandwidth (smaller is the value of $\chi$) beyond which ET-AIM is more efficient.

---

[2]The ET-AIM the iterations involve 3-D space FFTs to compute the immediate interactions, which scale as $O(N_\text{t} \bar{N}_\text{I} N_\text{C} \log N_\text{C})$; these FFT operations dominate other costs [7]. The TD-AIM iterations are devoid of any FFT operations and involve only sparse-matrix vector multiplications, which scale as $O(N_\text{t}^\text{TD} \bar{N}_\text{I}^\text{TD} N_0^\text{TD})$, where $N_0^\text{TD} \sim N_\text{s}$ [2],[7]. It is difficult to deduce how $\bar{N}_\text{I}^\text{TD}$ and $\bar{N}_\text{I}$ are related because they depend on the acceptable level of error in the iterative solution as well as the conditioning of the impedance matrices.

It is important to note again that the above theoretical deductions are of limited use because the assumptions listed at the beginning of Section 4.1 imply that the methods have different accuracy; specifically, ET-AIM incurs less integration errors [10] when computing the matrix entries and have smaller interpolations errors (see Section 3.1). A more meaningful comparison can be made empirically based on results that take into account the differences in the methods' accuracy-efficiency tradeoffs.

## 5 ET-AIM VS. TD-AIM EMPIRICAL COMPARISON

This section presents an empirical comparison of ET-AIM and TD-AIM based on optimized simulations of scattering from canonical objects that reside in free space. The methods' performances were compared while varying the object size and the bandwidth. The parameters in all the following simulations were optimized while ensuring that the error criterion in [7] was satisfied. This error criterion requires that the dB-error in the co-polarized range profile (the backscattered far-field envelope) is less than 2% when an 80-dB threshold is used similar to that in Section 2.4 (see (17) in [7]). The errors in the range profile, which include all frequencies of interest but only one scattering direction, are presented in [7] and the errors in co-polarized RCS using the measure in (2), which include all scattering directions but only one frequency, are presented in this paper. These errors were measured relative to the Mie series solution [18]; when this was impossible, a more accurate FD-AIM solution was used as reference [7].

Table I: Incident pulse parameters in benchmark simulations

| $\chi \approx 3$ | $\chi \approx 5$ | $\chi \approx 10$ | $\chi \approx 20$ | $\chi \approx 50$ | $\chi \approx 100$ |
|---|---|---|---|---|---|
| $f_c$=200 MHz | $f_c$=240 MHz | $f_c$=270 MHz | $f_c$=285 MHz | $f_c$=294 MHz | $f_c$=297 MHz |
| $\sigma = 3/200\pi$ µs | $\sigma = 1/40\pi$ µs | $\sigma = 1/20\pi$ µs | $\sigma = 1/10\pi$ µs | $\sigma = 1/4\pi$ µs | $\sigma = 1/2\pi$ µs |
| $t_d = 8\sigma$ | $t_d = 8\sigma$ | $t_d = 8\sigma$ | $t_d = 8\sigma$ | $t_d = 8\sigma$ | $t_d = 8\sigma$ |

### 5.1 Benchmark Scattering Problems and Optimized Parameters

Two different benchmark objects were simulated: A square plate and a sphere, which represent the best- and worst-case scenario for AIM, respectively. The side length of the plate $L_p$ was varied from $1$ m to $128$ m and the radius of the sphere $L_s$ was varied from $0.5$ m to $32$ m. In all the simulations that follow, the center frequency $f_c$ and standard deviation $\sigma$ of the excitation pulse were set such that $f_c + 3/2\pi\sigma$ was fixed to 300 MHz. The bandwidth was varied (Table I) by changing the $f_c$ and $\sigma$ from $f_c = 200$ MHz and $\sigma = 3/200\pi$ µs ($\chi \approx 3$) to $f_c = 297$ MHz and $\sigma = 1/2\pi$ µs ($\chi \approx 100$).

Table II: Spatial discretization for plate and sphere simulations

| Plate | $N_s = N_s^{TD}$ | TD | ET | Sphere | $N_s = N_s^{TD}$ | TD | ET |
|---|---|---|---|---|---|---|---|
| $L_p$:1 m | 280 | $14\times14\times1$ | $12\times12\times1$ | $L_s$:0.5 m | 684 | $18\times18\times18$ | $18\times18\times18$ |
| $L_p$:2 m | 1160 | $28\times28\times1$ | $20\times20\times1$ | $L_s$:1 m | 3384 | $27\times27\times27$ | $24\times24\times24$ |
| $L_p$:4 m | 4720 | $54\times54\times1$ | $36\times36\times1$ | $L_s$:2 m | 10 947 | $48\times48\times48$ | $48\times48\times48$ |
| $L_p$:8 m | 19 040 | $160\times160\times1$ | $72\times72\times1$ | $L_s$:4 m | 44 595 | $80\times80\times80$ | $64\times64\times64$ |
| $L_p$:16 m | 76 480 | $320\times320\times1$ | $144\times144\times1$ | $L_s$:8 m | 179 130 | $128\times128\times128$ | $128\times128\times128$ |
| $L_p$:32 m | 306 560 | - | $288\times288\times1$ | $L_s$:16 m | 742 059 | - | $256\times256\times256$ |
| $L_p$:64 m | 1 227 520 | - | $576\times576\times1$ | $L_s$:32 m | 2 903 916 | - | $512\times512\times512$ |
| $L_p$:128 m | 4 912 640 | - | $1152\times1152\times1$ | | | | |

The primary surface mesh, spatial basis functions, cubature rules, order of singularity extraction, CFIE linear combination parameter, and AIM matching scheme were set to be identical in ET-AIM and TD-AIM simulations; e.g., RWG basis functions [14] were used for spatial discretization, $\alpha = 1$ for plates and $\alpha = 0.5$ for spheres were used for the CFIE combination, and the moment matching method was used for computing anter/interpolation coefficients [2],[9],[19] such that moments up to order 3 were matched. The remaining parameters (auxiliary grid size, near-zone correction distance, temporal basis functions, oversampling rate, and iterative solver tolerance) were optimized independently for each method and benchmark simulation; it was observed that more accurate (and expensive) parameters are required for TD-AIM in order to achieve the requisite error level in the solution.

The auxiliary grid size for ET-AIM was chosen to be same as that in [13], except for the plates, where number of grid points in the $z$ direction was set to 1, which was found to be more efficient. The number of surface unknowns and grid sizes are listed in Table II.

For the sake of expediency, not all geometries were excited with all the different incident pulses; the simulations for which the parameters were optimized can be grouped into two sets: In the first set, the bandwidth was fixed ($\chi \approx 3$ and $\chi \approx 100$) and the problem size was varied; in the second set, the problem size was fixed ($N_s \in \{280, 19040\}$ for plate and $N_s \in \{684, 44595\}$ for sphere) and the bandwidth was varied.

Table III lists the essential time width, the temporal basis function type, and the time-step size used in these different simulations. As mentioned in Section 3.2, CPPIF of order $Q = 4$ was used for broader band TD-AIM simulations and $Q = 3$ was used for narrower band ET-AIM simulations; when the solutions were not accurate or stable, BLIF of half-width parameter $M = 5$ was used for the TD-AIM simulations and $M = 3$ was used for ET-AIM simulations. Table IV lists the near-zone correction threshold, the number of time steps needed to travel across the object, and the number of processes used in the simulations. As shown in Table IV, most of the simulations were performed using a single process but a few were run on many processes (up to $P=128$ for the largest sphere). The variable $\gamma$ in this table denotes the threshold distance normalized by the grid node spacing, beyond which $\tilde{\mathbf{Z}}_{l-l'}^{\text{FFT}}$ represents the scattered fields with the desired accuracy [2],[9]. The data in Table III shows that $\Delta t / \Delta t^{\text{TD}}$ was always greater than $\chi$; this is because the oversampling rate in ET-AIM simulations were always smaller than that in the corresponding TD-AIM ones, i.e., $\beta > \beta^{\text{TD}}$.

Table III: Temporal discretization for plate and sphere simulations

| | Temporal basis function type (either CPPIF of order $Q$ or BLIF of half-width $M$) Essential time-width ($\mu$s), Time-step size (ns) | | | | | | | | | | | | | | | | | |
|---|---|---|---|---|---|---|---|---|---|---|---|---|---|---|---|---|---|---|
| Bandwidth | $\chi \approx 3$ | | | $\chi \approx 5$ | | | $\chi \approx 10$ | | | $\chi \approx 20$ | | | $\chi \approx 50$ | | | $\chi \approx 100$ | | |
| Plate | $T^s$ | TD $Q=4$ | ET $M=3$ | $T^s$ | TD $Q=4$ | ET $M=3$ | $T^s$ | TD $Q=4$ | ET $M=3$ | $T^s$ | TD $Q=4$ | ET $Q=3$ | $T^s$ | TD $Q=4$ | ET $Q=3$ | $T^s$ | TD $Q=4$ | ET $Q=3$ |
| $L_p$:1 m | .08 | 0.19 | 2.0 | .11 | 0.19 | 3.3 | .22 | 0.19 | 8.3 | .38 | 0.19 | 16.7 | 1.0 | 0.19 | 27.8 | 1.90 | 0.19 | 47.6 |
| $L_p$:2 m | .07 | 0.19 | 2.0 | | - | | | - | | | - | | | - | | 2.00 | - | 47.6 |
| $L_p$:4 m | .13 | 0.28 | 2.0 | | - | | | - | | | - | | | - | | 2.09 | - | 47.6 |
| $L_p$:8 m | .18 | 0.24 | 1.67 | .22 | 0.19* (M=5) | 3.03 | .28 | 0.15* (M=5) | 5.6 | .57 | - | 9.5 | 1.2 | - | 23.8 | 2.21 | - | 41.7 |
| $L_p$:16 m | .22 | 0.24 | 1.67 | | - | | | - | | | - | | | - | | 2.38 | - | 47.6 |
| $L_p$:32 m | .27 | - | 1.67 | | - | | | - | | | - | | | - | | 2.38 | - | 47.6 |
| $L_p$:64 m | .28 | - | 1.67 | | - | | | - | | | - | | | - | | 2.38 | - | 47.6 |
| $L_p$:128 m | .31 | - | 2.0 | | - | | | - | | | - | | | - | | 2.38 | - | 47.6 |
| Sphere | $T^s$ | TD $Q=4$ | ET $M=3$ | $T^s$ | TD $Q=4$ | ET $M=3$ | $T^s$ | TD $Q=4$ | ET $M=3$ | $T^s$ | TD $Q=4$ | ET $Q=3$ | $T^s$ | TD $Q=4$ | ET $Q=3$ | $T^s$ | TD $M=5$ | ET $Q=3$ |
| $L_s$:0.5 m | .07 | 0.24 | 2.0 | 0.1 | 0.24 | 4.17 | 0.2 | 0.24 | 8.3 | 0.4 | 0.24 | 13.3 | 1.0 | 0.24 | 27.8 | 2.0 | 0.24 | 55.6 |
| $L_s$:1 m | .07 | 0.24 | 2.0 | | - | | | - | | | - | | | - | | 2.0 | 0.24 | 55.6 |
| $L_s$:2 m | .07 | 0.24 | 2.0 | | - | | | - | | | - | | | - | | 2.1 | 0.24 | 55.6 |
| $L_s$:4 m | .10 | 0.24 | 2.0 | 0.4 | 0.24 | 4.17 | 0.22 | 0.24 | 8.3 | 0.45 | 0.24* M=5 | 13.3 | 1.1 | 0.24* (M=5) | 33.3 | 2.3 | 0.24 | 55.6 |
| $L_s$:8 m | .13 | 0.21 | 2.0 | | - | | | - | | | - | | | - | | 2.5 | - | 55.6 |
| $L_s$:16 m | .20 | - | 2.0 | | - | | | - | | | - | | | - | | 2.5 | - | 55.6 |
| $L_s$:32 m | | - | | | - | | | - | | | - | | | - | | 2.7 | - | 66.7 |

A diagonally preconditioned GMRES iterative solver [20] was used for which the iterations are terminated when the error in the solution converges to a tolerance level, $\text{err}^{\text{tol}}$. The initial guess used by the iterative solver at each time step was set to the solution at the previous time step scaled by the ratio of the local tangential magnetic field at present time step to the previous one [7]. A higher iterative solver tolerance ($\text{err}^{\text{tol}} = 10^{-4}$) was used with ET-AIM than that for TD-AIM ($\text{err}^{\text{tol}} = 10^{-9}$) for all simulations, as it was observed to yield accurate enough results.

## 5.2 Computational Costs

Scattering of the various Gaussian pulses from the benchmark objects were computed using ET-AIM and when possible TD-AIM and ET-MOT. A similar study was performed in [7] using a different excitation pulse and a

Table IV: Additional parameters for plate and sphere simulations

| Bandwidth | χ≈3 TD | χ≈3 ET | χ≈5 TD | χ≈5 ET | χ≈10 TD | χ≈10 ET | χ≈20 TD | χ≈20 ET | χ≈50 TD | χ≈50 ET | χ≈100 TD | χ≈100 ET |
|---|---|---|---|---|---|---|---|---|---|---|---|---|
| **Plate** | | | | | | | | | | | | |
| $L_p$:1 m | 2 39 1 | 2 8 1 | 2 39 1 | 2 5 1 | 4 39 1 | 2 5 1 | 5 39 1 | 2 5 1 | 7 39 1 | 2 5 1 | 7 39 1 | 2 5 1 |
| $L_p$:2 m | 2 65 1 | 2 11 1 | - | - | - | - | - | - | - | - | - | 2 5 1 |
| $L_p$:4 m | 3 77 1 | 2 15 1 | - | - | - | - | - | - | - | - | - | 2 5 1 |
| $L_p$:8 m | 6 166 1 | 2 25 1 | 4 214 1 | 2 16 1 | 6 260 1 | 2 10 1 | - | 2 7 1 | - | 2 5 1 | - | 2 5 1 |
| $L_p$:16 m | 6 313 1 | 2 51 1 | - | - | - | - | - | - | - | - | - | 2 5 1 |
| $L_p$:32 m | - | 2 95 1 | - | - | - | - | - | - | - | - | - | 2 6 1 |
| $L_p$:64 m | - | 2 186 6 | - | - | - | - | - | - | - | - | - | 2 9 3 |
| $L_p$:128 m | - | 2 306 48 | - | - | - | - | - | - | - | - | - | 2 15 16 |
| **Sphere** | | | | | | | | | | | | |
| $L_s$:0.5 m | 3 30 1 | 3 7 1 | 3 30 1 | 3 5 1 | 3 30 1 | 3 5 1 | 3 30 1 | 3 5 1 | 3 30 1 | 3 5 1 | 3 30 1 | 3 5 1 |
| $L_s$:1 m | 3 44 1 | 3 9 1 | - | - | - | - | - | - | - | - | 3 44 1 | 3 5 1 |
| $L_s$:2 m | 3 72 1 | 3 12 1 | - | - | - | - | - | - | - | - | 3 72 1 | 3 5 1 |
| $L_s$:4 m | 3 129 2 | 3 19 1 | 3 129 2 | 3 5 1 | 3 129 2 | 3 5 1 | 3 129 2 | 3 5 1 | 3 129 2 | 3 5 1 | 3 129 2 | 3 5 1 |
| $L_s$:8 m | 3 274 32 | 3 33 8 | - | - | - | - | - | - | - | - | - | 3 5 1 |
| $L_s$:16 m | - | 3 59 64 | - | - | - | - | - | - | - | - | - | 3 5 16 |
| $L_s$:32 m | - | - | - | - | - | - | - | - | - | - | - | 3 6 128 |

Header spans: "Near-zone threshold ($\gamma$), Travel time across object ($N_g$), No. of processes ($P$)"

different error norm. The computational costs and the accuracy of the different simulations are shown as a function of the problem size and bandwidth in Figs. 2-3.

Fig. 2 shows the RHS time, iterative solution time, and total marching time for simulations performed for increasing size of plate $(280 \leq N_s \leq 4\,912\,640)$ and sphere $(684 \leq N_s \leq 2\,903\,916)$ and decreasing bandwidth of the excitation pulse $(3 \leq \chi \leq 100)$. The matrix-fill time and the memory requirement for these simulations can be found in [7]. The average number of iterations, the number of time steps, and the measured errors for these simulations are plotted in Fig. 3. In both figures, for the sake of clarity, the data obtained for fixed bandwidth (i.e., $\chi \approx 3$ and $\chi \approx 100$) and varying $N_s$ are also projected on the left wall and the data obtained for fixed problem size (i.e., $N_s = 280$ (684) and $N_s = 19040$ (44595) for plate (sphere)) and varying $\chi$ are also projected on the right wall.

Fig. 2 shows that the ET-MOT RHS time, iterative solution time, and total marching time all scale with the expected complexity of $O(N_s^2)$. Figs. 2(a)-(b) show that while the ET-AIM RHS time increases with problem size (in line with the complexity analysis), it initially decreases and later reaches a constant (for larger surfaces) as the bandwidth decreases. This is because, as the bandwidth narrows, the number of non-zero impedance matrices reduces until it reaches a constant at sufficiently large $\chi$, i.e., $N_g \sim \max(1, N_g^{\text{TD}}/\chi)$; e.g., the right wall in Fig. 2(a) shows that for the $N_s = 19040$ plate ( 44595 sphere), the ET-AIM RHS time for the $\chi \approx 100$ case is $\sim 3$ ($\sim 2$) times smaller than the $\chi \approx 3$ case. In fact, the larger the surface and therefore $N_g^{\text{TD}}$, the bigger the reduction in RHS time as the bandwidth narrows. Figs. 2(c)-(d) show that the ET-AIM iterative solution time increases with respect to the surface size and is essentially a constant with respect to relative bandwidth. The left wall in Fig. 2(c) shows that the ET-AIM iterative solution time for the plate increases faster ($\sim N_s \log^3 N_s$) than the expected ($\sim N_s \log N_s$) in the $\chi \approx 100$ case. This is because unlike the other cases where the average iteration count is about a constant, the iteration count increases as $\sim \log^2 N_s$ in this case (see Fig. 3(a)). Figs. 2(e)-(f) show that the ET-AIM marching time data is similar those observed in Figs. 2(c)-(d), i.e., the ET-AIM marching time is dominated by the iterative solution time.

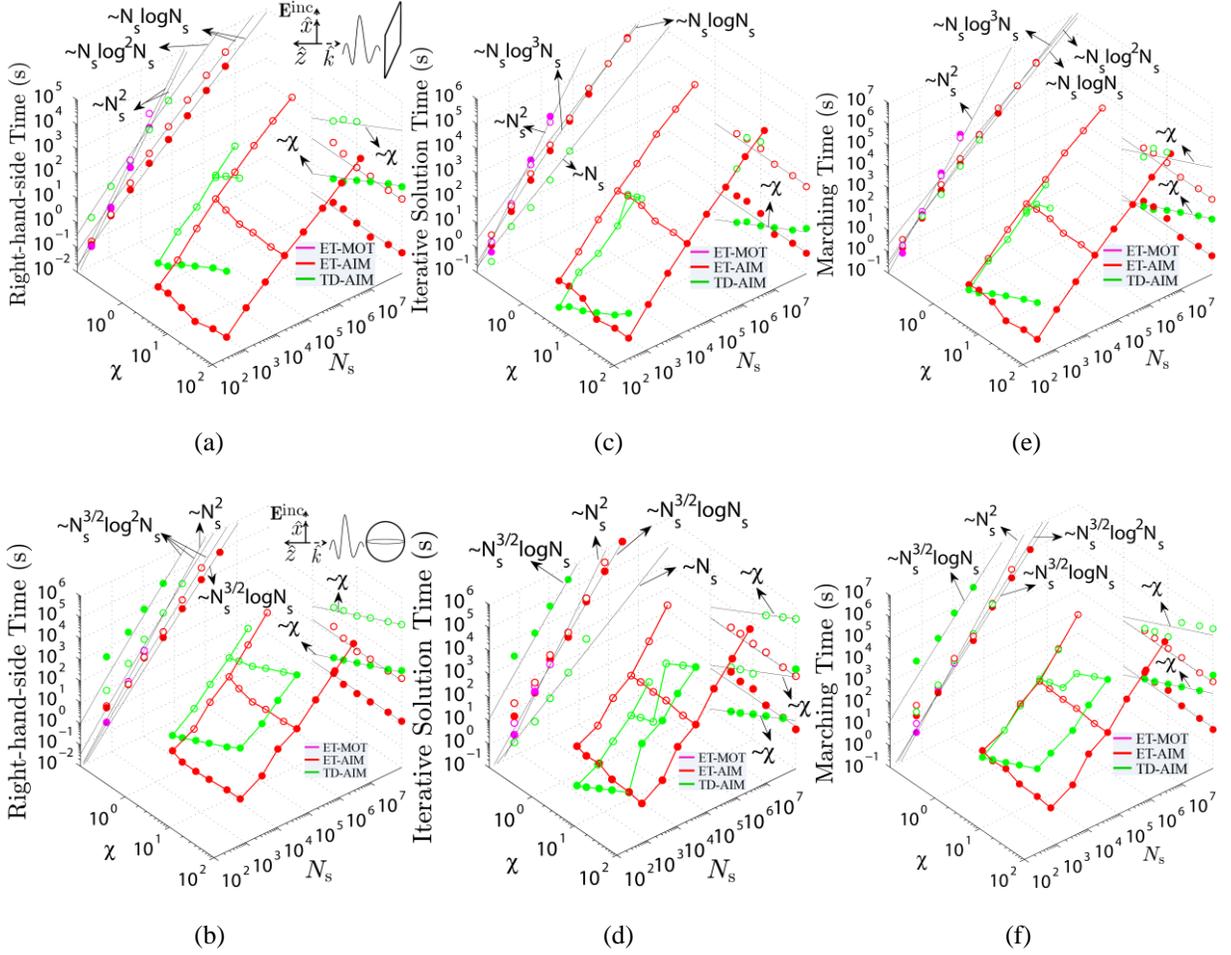

Figure 2: Scattering from a plate (top) and a sphere (bottom) as the surface area and the bandwidth is varied. Left: RHS time. Middle: Iterative solution time. Right: Marching time. All data on the side walls are identical to those in the 3-D plots except for the ET-MOT data, which is not shown in the 3-D plots for clarity. The left side walls show the parameters of interest as a function of $N_s$ for a narrow-band ($\chi \approx 100$, full symbols) and a broad-band ($\chi \approx 3$, empty symbols) case. The right walls show the same parameters as a function of $\chi$ for a small ($N_s = 280$ plate or $N_s = 684$ sphere, full symbols) and a large ($N_s = 19040$ plate or $N_s = 44595$ sphere, empty symbols) surface.

Figs. 3(a)-(b) show that the ET-AIM iteration count increases with the increase in problem size for narrow bandwidths especially when poorly conditioned EFIE was used; e.g., the left wall in Fig. 3(a) shows that the iteration count as $\sim \log^2 N_s$ for the $\chi \approx 100$ case. In contrast, the iteration count is essentially constant for the

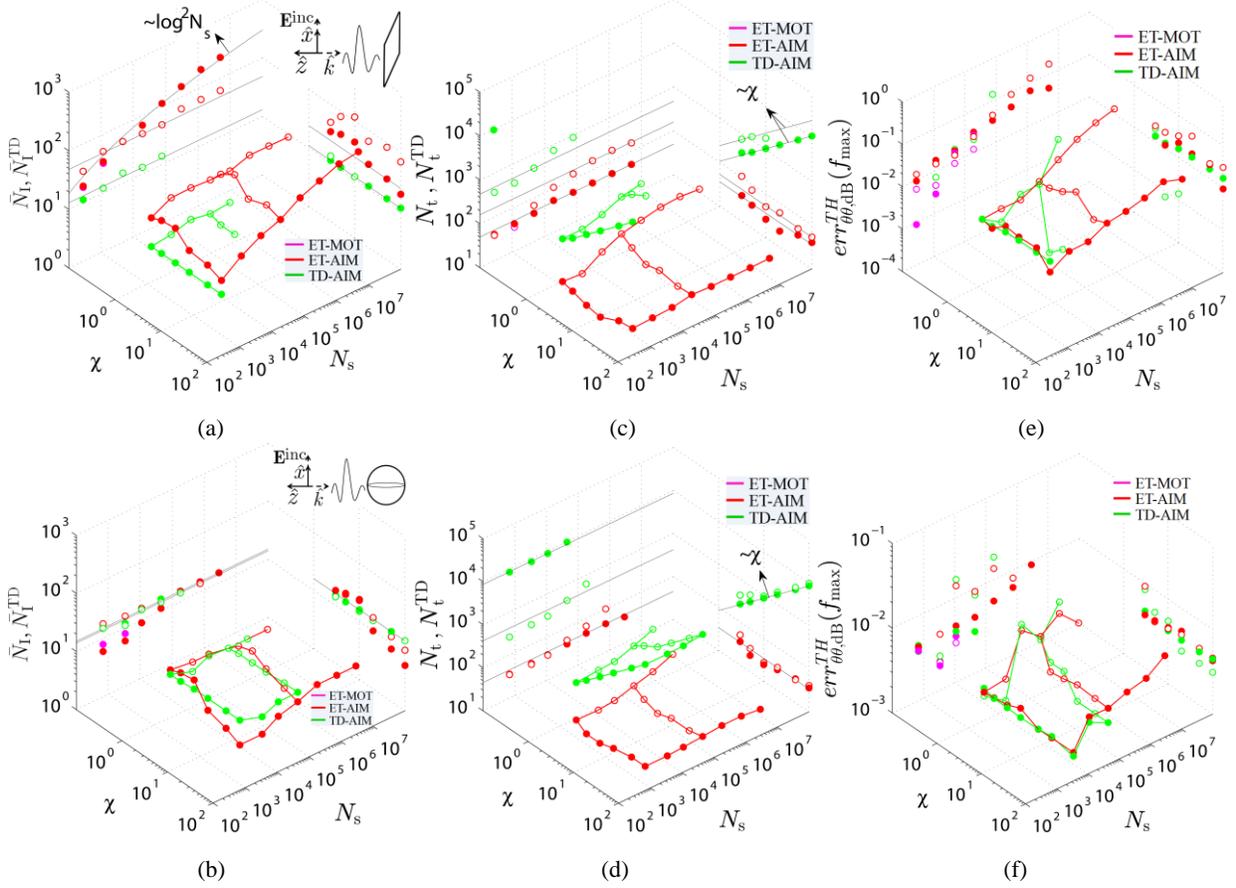

Figure 3: Scattering from a plate (top row) and a sphere (bottom row) as the surface size and the bandwidth is varied:: Left: Average number of iterations. Middle: Number of time steps. Right: Relative error in the bistatic RCS at $f_{\max}$. All data on the side walls are identical to those in the 3-D plots except for the ET-MOT data, which is not shown in the 3-D plots for clarity. The left side walls show the parameters of interest as a function of $N_{\rm s}$ for a narrow-band ($\chi \approx 100$, full symbols) and a broad-band ($\chi \approx 3$, empty symbols) case. The right walls show the same parameters as a function of $\chi$ for a small ($N_{\rm s} = 280$ plate or $N_{\rm s} = 684$ sphere, full symbols) and a large ($N_{\rm s} = 19040$ plate or $N_{\rm s} = 44595$ sphere, empty symbols) surface.

better-conditioned CFIE simulations in Fig. 3(b). Figs. 3(c)-(d) show that the number of time samples required for ET-AIM varies only weakly with problem size and the bandwidth. This is because for a given scattering object, $T^{\rm s}$ increased in proportion to $\chi$, i.e., the excitation pulse width dominated $T^{\rm s}$ and the scattered fields were not significant enough to extend the essential time width of interest. Figs. 3(e)-(f) shows that the ET-AIM RCS errors were less than $2\%$ for all the sphere simulations, but increased as the problem size increased for the plate simulations up to about $40\%$ for the largest plate in the $\chi \approx 3$ case. This is because with increasing plate size, the specular return or the main lobe in the RCS becomes stronger in magnitude and sharper in space, i.e., the RCS values are significantly smaller at angles away from the specular forward and backscattered directions. The inherent issue of ill conditioning related to EFIE limits the achievable accuracy for these relatively smaller values. Notice that these errors are exposed because the dB-error norm was used—a linear error norm would hide the errors in these smaller RCS values and result in errors less than $1\%$. It is also important to note that similar results and conclusions were obtained in [7] but using a different error norm as well as a different incident pulse shape (trapezoidal instead of Gaussian). ET-AIM is compared to the ET-MOT and TD-AIM next:

*ET-AIM vs. ET-MOT:* The ET-AIM marching was faster than the ET-MOT one for $N_s$ greater than $\sim 1000$ for the plate and $\sim 3000$ for the sphere (Figs. 2(e)-(f)). ET-AIM and ET-MOT iteration counts were close if not identical to each other (Figs. 3(a)-(b)). ET-MOT errors were lower than ET-AIM ones by about an order in magnitude for the narrow-band simulations for the first two plate sizes (Fig. 3(e)). ET-MOT errors were either comparable or slightly lower than ET-AIM ones for all other simulations (Fig. 3(e)-(f)).

*ET-AIM vs. TD-AIM:* As expected, the TD-AIM RHS time was always larger than the ET-AIM RHS time in all cases (Figs. 2(a)-(b)) and the TD-AIM iterative solution time was almost always smaller than the ET-AIM iterative solution time (Figs. 2(c)-(d))—the ET-AIM outperformed TD-AIM solution time only for very narrowband cases. As mentioned in the Introduction, the iterative solution time was more significant for ET-AIM whereas the RHS time was more significant for TD-AIM. This is because (i) ET-AIM has (much) cheaper RHS computations (primarily due to the reduction in the number of time steps and secondarily due to the reduction in $N_g$) and because (ii) ET-AIM has (slightly) more expensive iterative solution computations. The iterative solution was more expensive for ET-AIM either because of its higher cost per iteration (in the broadband cases) or because of its higher number of iterations (in the narrowband cases for the plate [Figs. 3(a)-(b)]), which somewhat offset the significant reduction in the number of time steps. It is important to highlight the large reduction in the number of time steps: $N_t^{TD}/N_t = \beta^{TD}\chi/\beta$ was $\approx 7-10$ in the plate simulations and $\approx 8-10$ in sphere simulations even for the broadest band $\chi \approx 3$ case (left wall, Figs. 3(c)-(d)). This ratio increased to $\approx 265$ in the plate simulations and $\approx 230$ in the sphere simulations for the narrowest band $\chi \approx 100$ case.

Overall, the ET-AIM marching time was comparable to TD-AIM in the broad-band simulations and was (significantly) faster as the bandwidth narrowed (Figs. 2(d)-(e)), e.g., the ET-AIM marching time for sphere simulations were $\approx 250$ times smaller than the corresponding TD-AIM ones for the narrow-band case ($\chi \approx 100$). .

ET-AIM and TD-AIM errors were found to be comparable (Figs. 3(e)-(f)).

## 6. COMPLEX SCATTERING PROBLEMS: MISSILE MODEL AND TRIHEDRON

The performance of ET-AIM when analyzing scattering from complex surfaces is illustrated in this section by simulating scattering from a missile model and a trihedron. The same geometries were also simulated in [7], where the errors were quantified by studying their range profiles. Here, additional validation is performed by comparing the RCS computed by ET-AIM to independent reference results. In the following simulations, unless specified otherwise, BLIFs with half-width parameter $M=3$ were used as the temporal basis function and the iterative solver tolerance was set as $\mathrm{err}^{\mathrm{tol}} = 10^{-4}$.

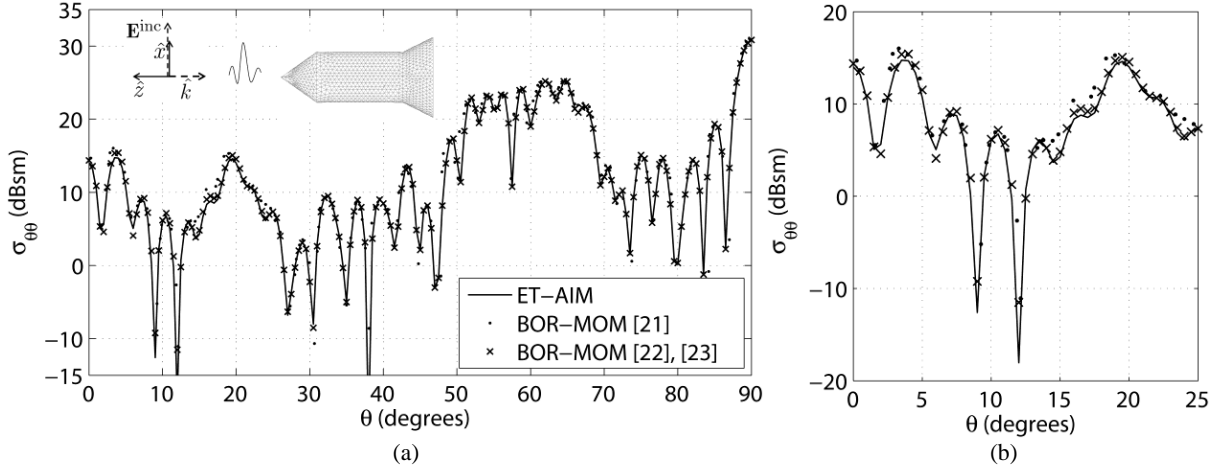

Figure 4: Monostatic RCS of the missile at $500$ MHz. ET-AIM results show discrepancies with respect to the independent BOR-MOM results in [21] especially in the 0 to 25 degree range, but agree well with BOR-MOM results computed by the authors using the formulation in [22],[23].

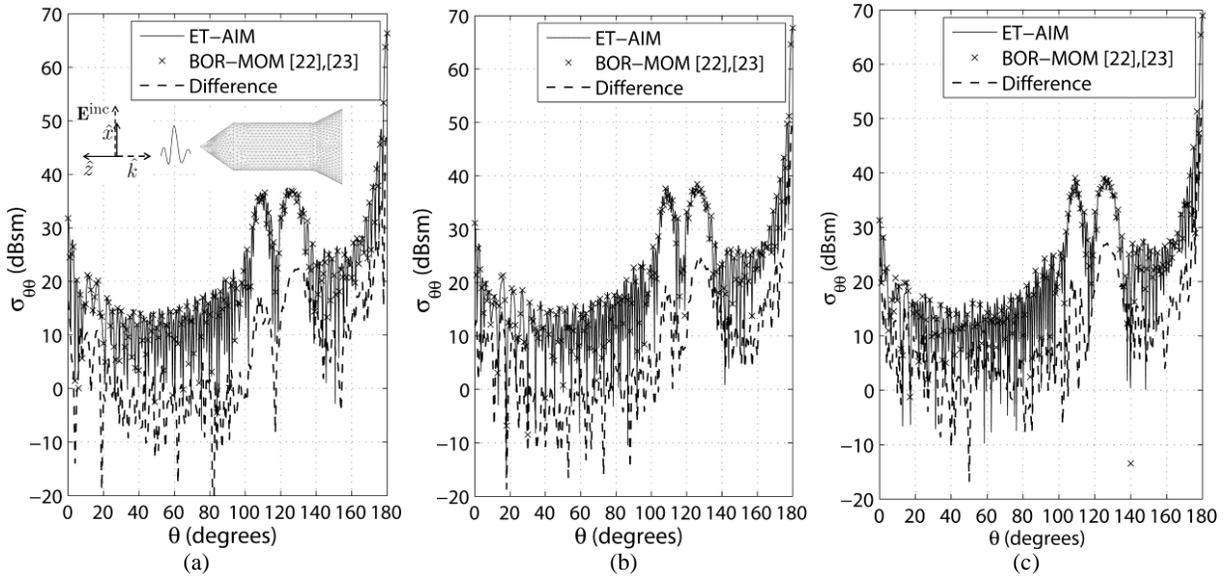

Figure 5: Bistatic RCS (VV) of the missile in the $x-z$ plane at frequencies: (a) $1.69$ GHz, (b) $1.82$ GHz, and (c) $1.95$ GHz.

First, the monostatic RCS of a PEC missile model residing in free space is compared to the reference results in [21]. The geometrical details of this model can also be found in [21]. The missile model was illuminated with the Gaussian plane wave in (1), where $\hat{p} = \hat{x}$, $\hat{k} = -\hat{z}$, $f_{\mathrm{c}} = 500$ MHz, $\sigma = 3/100\pi$ $\mu$s, and $t_{\mathrm{d}} = 8\sigma$. The essential bandwidth of the pulse is $f_{\mathrm{bw}} \approx 50$ MHz, i.e., $f_{\max} \approx 550$ MHz and $\chi \approx 11$. The surface current density was discretized using $N_{\mathrm{s}} = 166,080$ RWG basis functions, the auxiliary grid had $N_{\mathrm{C}} = 90 \times 90 \times 180$ nodes, and the near-zone parameter was set to $\gamma = 3$. The time-step size was $2$ ns resulting in $N_{\mathrm{g}} = 21$ and the essential time-width was $T^{\mathrm{s}} = 0.24$ $\mu$s ($N_{\mathrm{t}} = 120$). Fig. 4 shows the computed VV polarized monostatic RCS at

500 MHz, which exhibits reasonable agreement with the independent results in [21] obtained by a body-of-revolution method of moments (BOR-MOM) solver. To explain the discrepancy, the authors also developed a body-of-revolution method of moments (BOR-MOM) solver based on the formulation in [22],[23]; in this implementation, the number of harmonics in the Fourier series expansion to describe the variation of the current density in the azimuthal direction was set to $30$, the number of first order basis and testing functions to describe the variation in the transverse direction (along the generating curve) was $2050$, $50$- and $10$- point Gaussian quadrature rules were used for numerically computing integrals over the azimuthal and transverse direction, iterative solver tolerance was set to $10^{-6}$, and the CFIE formulation ($\alpha = 0.5$) was used. Figs. 4(a)-(b) show that the RCS found from ET-AIM agrees with the authors' BOR-MOM results at all angles; this indicates that the reference results in [21] were not converged at angles $\leq 25^{\circ}$.

Next, scattering from the same missile model is analyzed at a higher frequency: The excitation pulse was Gaussian plane wave as defined in (1), where $\hat{p} = \hat{x}$, $\hat{k} = -\hat{z}$, $f_c = 1.82$ GHz, $\sigma = 3/260\pi$ μs, and $t_d = 8\sigma$. The essential bandwidth of the pulse is $f_{bw} \approx 130$ MHz, i.e., $f_{max} \approx 1.95$ GHz and $\chi \approx 15$. The surface current density was discretized using $N_s = 2\,036\,019$ RWG basis functions, the auxiliary grid had $N_C = 270 \times 270 \times 540$ and the near-zone parameter $\gamma = 3$ was used for this simulation. CPPIF of order $Q = 3$ was used with the time-step size set to $\Delta t = 1.71$ ns, resulting in $N_g = 24$. The essential time-width was $T^s = 68.4$ ns ($N_t = 40$). The bistatic RCS was computed at different aspect angles in the $x - z$ plane. As shown in Figs. 5(a)-(c), good agreement is observed with the results using BOR-MOM [22],[23] at the minimum, center, and maximum frequencies. The measured error $\text{err}_{\theta\theta,\text{dB}}^{TH}$ at the minimum, center and maximum frequency were $2.75\%$, $1.60\%$, and $3.03\%$ for the corresponding error norm thresholds $TH$ of $-13.7$ dB, $-12.3$ dB, and $-11.1$ dB, respectively. The memory required for this simulation was $1.47$ TB. The matrix-fill, RHS, and solution times were about $88$ h, $8$ h, and $97$ h ($\bar{N}_I$ was $20$), respectively. It was not possible to obtain TD-AIM results because the corresponding simulation would require about 25 times more memory than the ET-AIM one.

Finally, ET-AIM is used to simulate scattering from a trihedron residing in free space [24]. The trihedron was illuminated using a Gaussian plane wave with $\hat{p} = -\hat{\theta}$, $\hat{k} = \hat{r}$, $f_c = 10$ GHz, $\sigma = 3/4\pi$ ns, and $t_d = 8\sigma$, where the unit vectors $\hat{\theta}$ and $\hat{r}$ are defined in the spherical coordinate system at polar angle $\theta = 100^{\circ}$ and azimuth angle $270^{\circ} \leq \phi \leq 360^{\circ}$. The essential bandwidth of the pulse is $f_{bw} \approx 2$ GHz, i.e., $f_{max} \approx 12$ GHz and $\chi \approx 6$. The surface current density was discretized using $N_s = 377,116$ RWG basis functions. The auxiliary grid had $N_C = 216 \times 480 \times 75$ nodes and near-zone grid parameter $\gamma = 3$. The time-step size was $87$ ps resulting in $N_g = 64$. The essential time-width was $T^s = 24.8$ ns ($N_t = 285$). Fig. 6 shows the computed monostatic RCS of the trihedron and compares it with the measurement data in [24] and independent simulation results from FISC

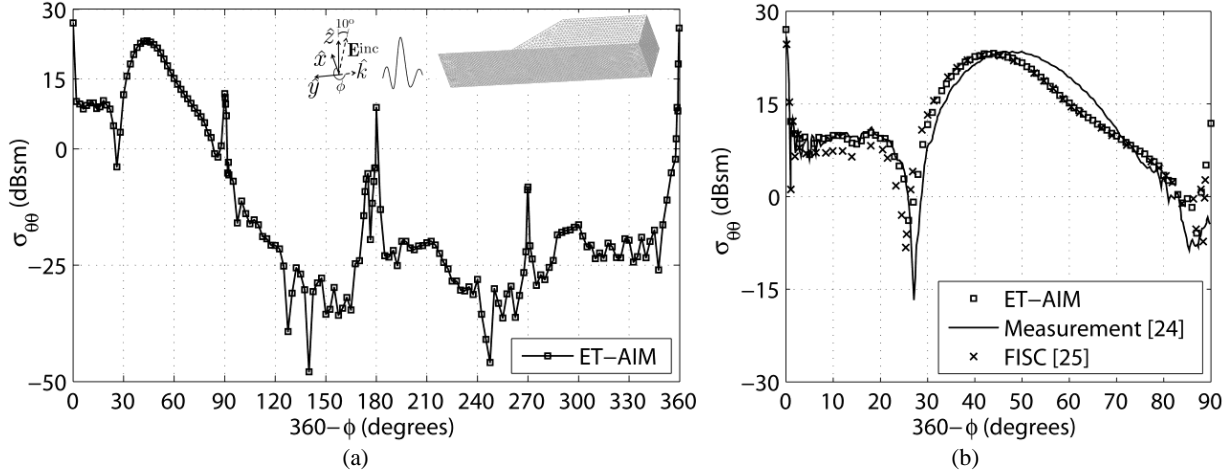

Figure 6: Monostatic RCS of the trihedron at 10 GHz for (a) entire range of azimuthal angle (b) $270° \leq \phi \leq 360°$.

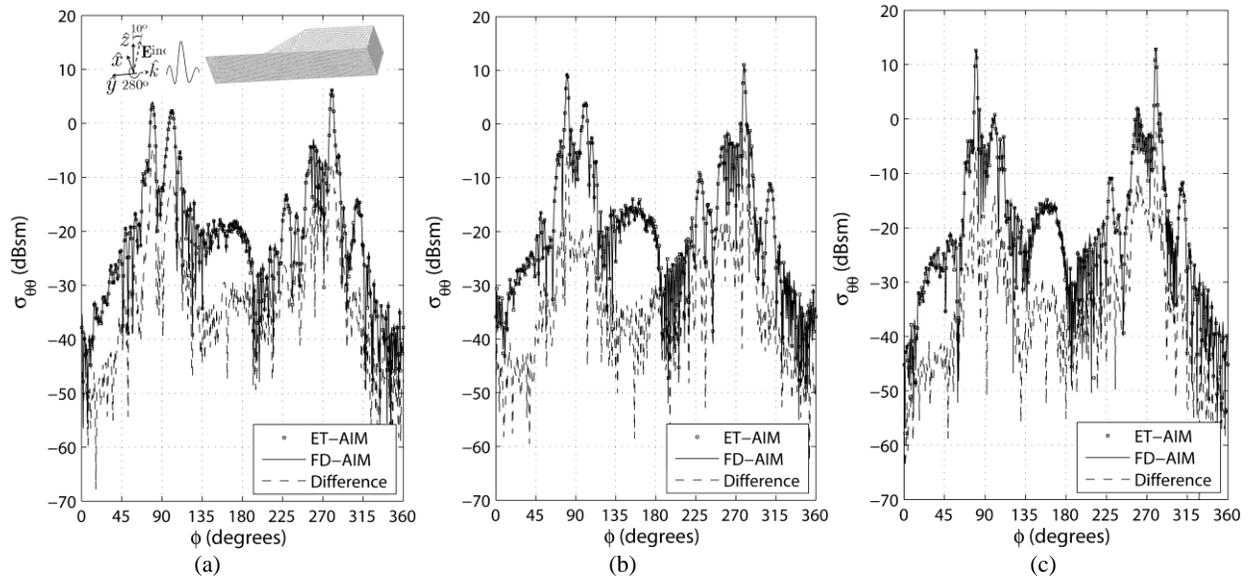

Figure 7. Bistatic RCS of the trihedron for $\theta=80°$ at (a) 8 GHz, (b) 10 GHz, and (c) 12 GHz.

[25], also provided in [24]. The ET-AIM results in Fig. 6(b) show reasonable agreement with the measured data and the discrepancies for $360° - \phi > 20°$ are similar to those observed in FISC results; thus, the disagreement is likely due to the differences between the modeled and measured structures rather than due to errors in the numerical solution method. The memory required for this simulation was $\sim 417$ GB. The matrix-fill, RHS, and solution times for each incident angle were about $4.5$ h, $6.9$ h, and $263$ h ($\bar{N}_\mathrm{I}$ was $67$) respectively. The TD-AIM simulation could not be performed as a result of high memory requirement, estimated to be $3.6$ TB. For further validation, bistatic RCS is computed for the pulse incident from the $\theta = 100°$ and $\phi=280°$ direction. The result is plotted with respect to azimuthal angle for $\theta = 80°$ and shows good agreement with the reference frequency-domain result (Fig. 7).

# 7 CONCLUSIONS

This paper presented a detailed empirical analysis of ET-AIM and compared it to its time-domain counterpart, TD-AIM. ET-AIM was observed to yield more accurate results than TD-AIM when using identical parameters. As a result, parameters such as auxiliary grid size, near-zone correction distance, temporal basis functions, oversampling rate, and iterative solver tolerance were optimized independently for the two methods, taking their different accuracy-efficiency tradeoffs into account. Empirical comparisons thus obtained show two key results in the high-frequency regime of analysis: (i) The iterative solution time is the dominant component of ET-AIM marching cost whereas the right-hand-side time is dominant for TD-AIM. (ii) ET-AIM has comparable marching cost to TD-AIM for the broadest band case ($\chi \approx 3$) but as the bandwidth narrows, ET-AIM becomes $\sim \chi$ times faster than TD-AIM and is more stable for large surfaces. Moreover, the algorithm always has lower memory requirement than TD-AIM and the reduction becomes more significant as the bandwidth narrows [7]. As a result, ET-AIM enables the solution of larger and more complex transient scattering problems that are impractical to obtain using TD-AIM.

The results in this paper and [7] show that ET-AIM should be preferred over TD-AIM for transient scattering analysis in all the cases considered; they also imply that envelope-tracking integral-equation methods should outperform time-domain ones in general. A possible exception is when solving broadband scattering problems for which the iterative solution times for envelope-tracking methods are significantly more expensive (due to ill conditioning, resonances, etc.) while the RHS times for time-domain methods remain comparable to the cases considered here—a possibility when the realism of the object model increases and the scattering problem cannot be classified as a high- or low-frequency problem. This is also a well-known shortcoming of frequency-domain integral-equation methods compared to time-domain ones; in fact, envelope-tracking methods trade off the cheaper iterative solution of time-domain methods in return for a significant reduction in the number of time steps in the analysis, the memory requirement, and the right-hand-side time. The results in [7] indicate that iterative solutions continue to converge faster for envelope-tracking methods compared to frequency-domain methods and the gains relative to time-domain methods are well worth any increases in iterative solution time when analyzing scattering from spheres, plates, a trihedron, a missile model, and a model aircraft. Whether the tradeoff remains advantageous for envelope-tracking methods when solving even more complex scattering problems remains to be seen.